\documentclass[aps,twocolumn]{revtex4}
\usepackage{amsfonts,amssymb,amsmath}            
\usepackage[]{graphics,graphicx,epsfig} 
\usepackage{graphicx}
\usepackage{dcolumn}
\usepackage{bm}
\usepackage{amsthm}

\newtheorem{lemma}{Lemma}

\newtheorem*{definition}{Definition}

\newtheorem{thm}{Theorem}

\def\vec#1{{\bf#1}}

\def\identity{\leavevmode\hbox{\small1\kern-3.8pt\normalsize1}}
\def\ket#1{| #1 \rangle}
\def\bra#1{\langle #1 |}
\newcommand{\braket}[2]{\left\langle #1|#2\right\rangle}
\newcommand{\proj}[1]{\ket{#1}\bra{#1}}
\def\ip#1#2{\langle #1 | #2 \rangle}
\def\braket#1#2{\langle #1 | #2 \rangle}

\def\norm#1{\| #1 \|}

\def\dim{\operatorname{dim}}

\def\Tr{\operatorname{Tr}}

\def\H{\mathcal{H}}

\def\uu{\mathfrak{u}}
\def\su{\mathfrak{su}}

\def\UU{\mathfrak{U}}
\def\sp{\mathfrak{sp}}

\def\so{\mathfrak{so}}

\def\targ{{\rm targ}}
\def\In{{\rm in}}
\def\out{{\rm out}}

\newcommand{\half}{\mbox{$\textstyle \frac{1}{2}$}}
\renewcommand{\epsilon}{\varepsilon}
\begin{document}

\title{Quantum Control Theory for State Transformations: \\ Dark States and their Enlightenment}
\date{\today}

\author{Peter J.~\surname{Pemberton-Ross}}
\author{Alastair~\surname{Kay}}
\author{Sophie G.~\surname{Schirmer}}
\affiliation{Centre for Quantum Computation,
             Dept of Applied Maths and Theoretical Physics,
             University of Cambridge,
             Wilberforce Road,
             Cambridge CB3 0WA, United Kingdom}

\begin{abstract}
For many quantum information protocols such as state transfer,
entanglement transfer and entanglement generation, standard notions of
controllability for quantum systems are too strong.  We introduce the
weaker notion of accessible pairs, and prove an upper bound on the
achievable fidelity of a transformation between a pair of states based
on the symmetries of the system.  A large class of spin networks is
presented for which this bound can be saturated. In this context, we
show how the inaccessible dark states for a given excitation-preserving
evolution can be calculated, and illustrate how some of these can be
accessed using extra catalytic excitations. This emphasizes that it is
not sufficient for analyses of state transfer in spin networks to
restrict to the single excitation subspace. One class of symmetries in
these spin networks is exactly characterized in terms of the underlying
graph properties.
\end{abstract}
\maketitle

\section{Introduction}

The holy grail of quantum information is the realization of a universal
quantum computer, capable of solving a variety of tasks, including some
that cannot be solved efficiently by classical computers.  Although
progress has been made towards this goal, it remains a daunting task.  A
more practical alternative is to develop utility modules capable of
specific tasks such as
quantum wires for state transfer \cite{christandl1,akreview} or entanglement generation \cite{kay:ghz,kwek,akreview}, which are easier to design and manufacture.  These
tasks can be achieved with minimal or no control.  Perfect state
transfer, for example, can be achieved by propagation along a spin chain
with a suitably engineered Hamiltonian~\cite{christandl1,akreview}.
However, these schemes have obvious drawbacks such as the requirement for
precise Hamiltonian engineering, and the fact that they are limited to
performing a single fixed task.  We would like to relax the constraints
slightly and see how much is gained by adding a small amount of control.
Do we enlarge the range of tasks that can be achieved and the systems
that can be used?

Recent work suggests that it is indeed possible to vastly increase the
potential of many systems with very limited control. Control of a single
coupling and some local control of the end spins, for example, can be
sufficient to efficiently realize a universal quantum computer
\cite{ak_and_me,burgarth_other}.  The problem of realizing quantum information
primitives such as state transfer and entanglement generation using a
minimal degree of control, and the related task of identifying the key
properties of a system such that those control techniques can then be
applied, have also been studied, in particular for spin systems.  Most
studies have focused on special geometries, particularly chains and
quasi-one-dimensional systems \cite{kay:ghz,kwek,ak_and_me,burgarthkojinori, bosereview,
christandl1, akreview,rap_comm_2009}, dual-chains \cite{burgarthdual}
and rings \cite{bosering}, but there have been some studies more general
networks~\cite{lloyd_sev,simsev1}.  Each of these investigations has
proved a certain information processing property, but a general
framework for determining the capability of an arbitrary network is
still lacking.

In this paper, we propose a unifying framework for describing certain
state transformation tasks that involve simultaneously transforming a
fixed set of initial states into a fixed set of output states.  This
allows us to simultaneously discuss processing tasks such as perfect
state transfer and entanglement generation, rather than tackle each case
separately.  These tasks require far less than full control of
the system, but the ability to perform them is restricted by symmetries
of the system.  In Sec.~\ref{sec:2} we discuss the notions of full
controllability, and introduce the much weaker notion of sets of
accessible pairs, and describe the role of Lie algebra symmetries,
deriving bounds on the fidelity with which a given state transformation
task can be realized based on symmetry properties of the system and
control Hamiltonians.  In Sec.~\ref{sec:3} we demonstrate how the upper
bounds on a state transformation task can be achieved within a large
class of spin networks by controlling a single coupling between two
spins. This includes an exact description of one class of symmetries for
these systems, a necessary and sufficient condition for their presence
being that the underlying network topology should correspond to a
bipartite (two-colorable) graph.  In Sec.~\ref{sec:dark_states} it is
shown how symmetry restrictions can be overcome by catalytic
excitations, which emphasizes a marked difference in state
transformation capabilities in comparison to the standard assumptions
made in the analysis of state transfer in networks \footnote{For the
initial studies which considered chains, this assumption was well
motivated because there is no catalysis in this case.}. Finally, in
Sec.~\ref{sec:5}, we discuss how many of the important parameters for
the transfer can be estimated from experimental data if they are not
previously known.

\section{Symmetries \& Accessible States} 
\label{sec:2}

\subsection{Controllability} 

Among the most elementary concepts in control theory are reachability
and controllability. Given a controlled dynamical system, a state
$\vec{x}_1$ is {\em reachable} from an initial state $\vec{x}_0$ if
there exists a control sequence and some time $T$ such that the
corresponding trajectory of the system satisfies $\vec{x}(0)= \vec{x}_0$
and $\vec{x}(T)=\vec{x}_1$, and the system is {\em controllable} if any
state in the state space is reachable from any other state \footnote{The
term {\em globally controllable} is often coined here, as in
\cite{ddabook}.  However, we choose the alternative term to enable
certain distinctions with, for instance, the subject that has come to be known as global control \cite{global1}
that one of the authors has worked on \cite{akglobal}, and the fact that
full control can be realized with a single local coupling
\cite{ak_and_me}.}.  Applied to bilinear control systems evolving on a
Lie group, one can derive explicit criteria for controllability in terms
of the Lie algebra formed by the dynamical generators.  In particular,
this is the case for decoherence-free quantum systems governed by the
Schr\"{o}dinger equation $\dot{U}(t)=-iH(t)U(t)$, where $U(t)$ is a
unitary operator on the Hilbert space $\H$ of the system, and $H(t)$ is
a Hamiltonian that depends linearly on a set of (real-valued) controls
$f_m(t)$
\begin{equation}
  \label{eqn:generic_ham}
   H(t) = H_0 + \sum_{m} H_m f_m(t).
\end{equation}
Here $H_0$ is the system Hamiltonian and the $H_m$ are Hamiltonians
governing the interaction with time-varying control fields $f_m(t)$.
$iH_0$ and $iH_m$ are the generators of the dynamics and they generate
the {\em dynamical Lie algebra} $L$, which consists of all linear
combinations and iterated commutators of these generators.  For a system
with Hilbert space dimension $n$ the generators $iH_m$, $m=0,1,\ldots$,
are anti-Hermitian matrices and thus can, at most, generate a subalgebra
of the Lie algebra $\uu(n)$ of anti-Hermitian $n\times n$ matrices.  A
necessary and sufficient condition for the
system~(\ref{eqn:generic_ham}) to be fully controllable then is the Lie
algebra rank condition~\cite{ddabook}, $L\simeq \uu(n)$.  If it is
satisfied, then it can be shown that for any sufficiently large $T$,
there exists a set of controls $f_m(t)$ such that any desired unitary
gate $U_{\targ}\in\UU(n)$ can be realized, i.e., given any $U_{\targ}
\in\UU(N)$, we have~\cite{Sussman72}
\begin{equation}
  U_{\targ} = U(T) =  
  \exp_{+}\left\{-i \int_{t=0}^T\!\!\! H_0 + \sum_m f_m(t) H_m \, dt\right\},
\end{equation}
where $\exp_+$ denotes a positive time-ordered exponential.  The
reachable set $K=\exp(L)$, known as the {\em dynamical Lie group} is, in
this case, the entire unitary group $\UU(n)$.  Applied to an $N$-qubit
system with a control-linear Hamiltonian~(\ref{eqn:generic_ham}), this
implies that the system is fully controllable if and only if
$L=\uu(2^N)$, and the corresponding dynamical Lie group $K=\UU(2^{N})$.
In practice, this condition can be slightly relaxed as we can usually
neglect the global phase of the system, in which case $L=\su(n)$
suffices, where $\su(n)$ is the Lie algebra of trace-zero anti-Hermitian
$n\times n$ matrices.  In this paper we consider systems which are not
fully controllable, for which $L$ need not be $\uu(n)$ or $\su(n)$.

\subsection{Sets of Accessible Pairs}

The main task in which we are interested throughout this paper is
whether a system described by Eqn.~(\ref{eqn:generic_ham}) can be caused
to transform its initial state $\ket{\psi_{\text{in}}}$ into a given
target state $\ket{\psi_{\targ}}$ through suitable manipulations of the
control fields $f_m(t)$.  If this is the case, we say
$(\ket{\psi_{\text{in}}},\ket{\psi_{\targ}})$ is an \emph{accessible
pair} of the system. If a given pair is not accessible then we quantify
how well the system can achieve the desired transformation by
calculating the fidelity of the optimal output state
$\ket{\psi_{\text{out}}}$,
\begin{equation*}
  F:=|\braket{\psi_{\text{out}}}{\psi_{\targ}}|^2. 
\end{equation*} 
Accessible pairs have $F=1$, whereas a state is dark with respect to a
given input if $F=0$.  One can extend this concept to sets of accessible
pairs.  We say a system has a set of (simultaneously) accessible pairs
$\left\{(\ket{\psi_{\text{in}}^1},\ket{\psi_{\targ}^1}),
(\ket{\psi_{\text{in}}^2},\ket{\psi_{\targ}^2})\right\}$ if the state
$\ket{\psi_{\text{in}}^1}$ can be transformed into
$\ket{\psi_{\targ}^1}$ by some set of manipulations, and the same set of
manipulations, in the same time, also transforms
$\ket{\psi_{\text{in}}^2}$ into $\ket{\psi_{\targ}^2}$.

This formalism provides a simple way for describing state transfer.
Consider two nodes of a network $A$ and $B$. State transfer from $A$ to
$B$ can be achieved if
\begin{equation*}
  \left\{(\ket{0}_A\ket{\phi},\ket{\psi}\ket{0}_B),
         (\ket{1}_A\ket{\phi},\ket{\psi}\ket{1}_B)
  \right\}
\end{equation*}
are accessible pairs for any arbitrary states $\ket{\phi}$ and
$\ket{\psi}$ as it implies
\begin{equation*}
  (\alpha\ket{0}_A+\beta\ket{1}_A)\ket{\phi} \mapsto
   \ket{\psi}(\alpha\ket{0}_B+\beta\ket{1}_B)
\end{equation*} 
for any superposition $\alpha\ket{0}+\beta\ket{1}$ (In this paper, we will only consider the decoherence-free case, which permits us to use linearity of the transformation task).  Previous studies
have focused almost exclusively on systems with excitation-preserving
Hamiltonians because this vastly simplifies the task.  As
$\ket{0}^{\otimes N}$ is an eigenstate of such a system, setting
$\ket{\phi}=\ket{\psi}=\ket{0}^{\otimes N-1}$ shows that we simply
require that the system has an accessible pair
$(\ket{1}_A\ket{0}^{\otimes N-1},\ket{0}^{\otimes N-1}\ket{1}_B)$.  We
also mention that if we have an initial state
$\frac{1}{\sqrt{2}}(\ket{00}+\ket{11})_{A,A'}\ket{\phi}$, where $A'$ is
an ancilla qubit not part of the system, then the same accessible set as
for state transfer allows the entanglement to be distributed between
$A'$ and $B$, $\frac{1}{\sqrt{2}}(\ket{00}+\ket{11})_{B,A'}\ket{\psi}$.

The concept of accessible pairs is a very weak property compared to
full controllability.  It requires only that we can implement \emph{one}
unitary operator that simultaneously maps the input states
$\ket{\psi_{\In}^{k}}$ to the target states $\ket{\psi_{\targ}^{k}}$, as
opposed to an \emph{arbitrary} unitary operator.  In the example above,
this means we only need to generate one unitary operator
$U$ that maps $\ket{10\cdots0}$ to $\ket{00\cdots01}$ to swap the
quantum states of the first and last qubit. A one-dimensional Lie
algebra may suffice for these purposes \cite{christandl1}. 

\subsection{Reducibility and commuting symmetries}

The most important reason for lack of full controllability of a control
system on a Lie group are symmetries restricting the dynamical Lie
algebra and dynamical Lie group.  The relation between symmetries in the
Hamiltonian of the system, irreducibility and loss of full control has
been studied in recent papers by Polack \textit{et al}.~\cite{tannor}
and Sander \textit{et al}.~\cite{tosh}.  Lie algebra symmetries can be
classified in various ways.  We shall distinguish between symmetries
that imply reducibility of the dynamical Lie algebra and other symmetries. 

A dynamic Lie algebra $L$ is said to be {\em reducible} if it can be
written as the direct sum of simple Lie algebras, e.g., $L \simeq \oplus_{i} L_{i}$. A necessary and sufficient condition for reducibility of
$L$ is the existence of a non-trivial Hermitian symmetry operator $J$ that
commutes with every element of $L$,
\begin{equation*}
  [H_m,J] =0, \quad \forall\: m=0,1,2,\ldots.
\end{equation*}
We shall refer to such $J$ as {\em commuting symmetry operators} (CSOs).
The commutation relations imply that the Hamiltonians leave non-trivial
subspaces of the Hilbert space $\H$ invariant, and thus there exists a
basis with respect to which $J$ and all $H_m$ have the same non-trivial
block-diagonal structure, and we can write the Hilbert space $\H$ as a
direct sum of mutually invariant subspaces $\H=\oplus_{d=1}^D \H_d$,
where $K$ is the number of independent blocks.  Every $H_m$ has an
eigenvector decomposition such that every eigenvector 
can be written as a linear superposition of the
eigenvectors of $J$ drawn from a given degenerate eigenspace.  The
evolution on each subspace is described by its own Hamiltonian and thus
the projection of a state onto each subspace, $\norm{\Pi_d\psi(t)}$, is
a constant of motion.  Therefore, a necessary condition for a system to
have an accessible pair $(\ket{\psi_{\text{in}}},\ket{\psi_{\targ}})$ is
that the initial state $\ket{\psi_{\text{in}}}$ and target state
$\ket{\psi_{\targ}}$ have the same projections onto each invariant
subspace.

An important subclass of CSOs are the \emph{permutation
symmetries}, corresponding to the interchange of indistinguishable
physical qubits of the $N$-qubit system.  This particular subclass was
referred to as an outer symmetry in \cite{tosh}.  A general procedure to
find CSO is described in Appendix~\ref{appendix:A}.

\subsection{Irreducibility and other symmetries}

Irreducibility of the Lie algebra on an $n$-dimensional subspace $\H_k$
alone, however, is insufficient to ensure full controllability on this
subspace.  Assuming the Hamiltonians $H_m$ are traceless, which can be
achieved by subtracting a multiple of the identity if necessary, all
irreducible components of the Lie algebra must be simple subalgebras of
$\su(n)$, which restricts to candidates of the form
$\sp(k)$, $\so(k)$, $\su(k)$ (for some $k<n$) or one of the exceptional
Lie algebras.  In this case, although we can always produce a unitary
evolution that takes an initial state to a final state with a non-zero
overlap with our target state, we may not be able to produce arbitrarily
high overlap with the target~\cite{tannor}.  For a subspace of dimension
$n$ this occurs precisely when the dynamical Lie algebra is neither
$\su(n)$ nor $\sp(\frac{n}{2})$~\cite{albertini}.  

\textbf{Example 1:} A simple example is a system with
\begin{equation*}
  H_0 = \frac{\omega}{2}
      \begin{pmatrix}
       -1 & 0 & 0 \\ 0 & 0 & 0 \\ 0 & 0 & 1
      \end{pmatrix}, \quad
  H_1 = \begin{pmatrix} 0 & d & 0 \\ d & 0 & d \\ 0 & d & 0 \end{pmatrix}.
\end{equation*}
The Lie algebra generated by $\{iH_0, iH_1\}$ is a unitary
representation of $\so(3)$ and indecomposable, but the system possesses
an orthogonal symmetry; both matrices anticommute with
$$
\left(\begin{array}{ccc}
0 & 0 & 1 \\
0 & -1 & 0 \\
1 & 0 & 1
\end{array}\right).
$$
Let us define the basis to be $\ket{\pm\lambda}$ and $\ket{1}$, where $\ket{\pm\lambda}$ are the eigenvectors of $H_0$ with eigenvalues $\pm1$ respectively. $H_1$ couples the state $\ket{1}$ equally to the states $\ket{\pm\lambda}$, which are not directly coupled. Imagine we start from the state $\ket{1}$ and wish to produce a state $\ket{\Psi}$. The symmetry prevents us from perfectly producing $\ket{\Psi}$ unless 
$$
|\braket{\lambda}{\Psi}|=|\braket{-\lambda}{\Psi}|.
$$
To see this, consider the unitary
basis change
\begin{equation*}
  B = \frac{1}{2}
       \begin{pmatrix} 
        1 & \sqrt{2}i & 1 \\
        \sqrt{2}i & 0 & -\sqrt{2}i \\
        1 & -\sqrt{2}i &1 
       \end{pmatrix},
\end{equation*}
which maps the generators $iH_0$ and $iH_1$ onto real skew-symmetric
matrices. Hence the evolution under any control sequence
$$
B\:\text{exp}_+\left\{-i\int_{t=0}^TH_0+f(t)H_1dt\right\}B^\dagger
$$
must be described by a unitary matrix which is real. Since $B\ket{1}$ is a real vector (up to a global phase), the output $B\ket{\Psi}$ must be a real vector $(a,b,c)^T$. Hence, inverting $B$ shows that we have $\braket{\pm\lambda}{\Psi}=\half(a+c\mp\sqrt{2}ib)$ which, because each of the coefficients is real, have equal magnitudes of amplitude.

Thus, given a maximal decomposition of the Hilbert space and Lie algebra
into irreducible subalgebras, the next step is to identify symmetries
such as orthogonal or symplectic symmetries, which are characterized by
the existence of an orthogonal or symplectic symmetry operator $J$ such that every element $x$ in
the irreducible Lie algebra $L_{i}$ satisfies 
\begin{equation}
 \label{eqn:antisymmfull}
 x^T J + J x=0.
\end{equation} 
If all
the dynamic generators of the system are of the form $x=iH_m$, where
$H_m$ are real-symmetric matrices, then $(iH_m)^T J + J (iH_m)=0$
simplifies to the anticommutation relation 
\begin{equation}
 \label{eqn:antisymmsimple}
\{H_m,J\}=0  \quad \forall m=0,1,2,\ldots
\end{equation}
 and we shall therefore refer to such $J$ as \emph{anticommuting symmetry operators}
(ASOs).  Given the Hamiltonians $H_m$, such symmetries can easily be
determined, as described in Appendix~\ref{appendix:A}.

\subsubsection{Simultaneous Symmetries}

A given system may have several CSOs that do not commute with
each other, meaning that block-diagonal structures are not
simultaneously realizable. We will always choose to analyze the symmetries that are most natural for the problem at hand, but one might worry that other non-commuting symmetries, be they CSOs or ASOs, could come into play. We will briefly justify that they do not, and therefore once we have picked a set of CSOs, we only have to find other CSOs or ASOs that act on the reduced space.

Imagine that we have found a CSO such that the system Hamiltonian and the control Hamiltonians all decompose into the same subspace structure,
$$
\mathcal{H}=\mathcal{H}_1\oplus\mathcal{H}_2
$$
and that our initial state only has support on $\mathcal{H}_1$. Clearly, the evolution of that initial state is only determined by Hamiltonians restricted to $\mathcal{H}_1$; it does not matter what the Hamiltonians are on the complement to that space. So, let's say we have now found an $M$ which could either be a CSO or ASO, but does not respect the subspace structure i.e.~it couples between the two. The existence of this $M$ must depend heavily on the structure of the Hamiltonians on the complement space -- if they were different, $M$ would not exist, and then it certainly couldn't restrict the dynamics within $\mathcal{H}_1$. Since we've argued, however, that our evolution of a state supported on $\mathcal{H}_1$ should not be affected by changing the Hamiltonians on the complement space, it must be that $M$ does not affect the dynamics.

\subsection{Maximum information transfer fidelity}

Combining the knowledge of how the system Hamiltonian decomposes under
basis change with the knowledge of the symmetries satisfied by each
subsystem allows us to characterize what properties of the system's
population are conserved under controlled evolution.  This information
can be used to calculate the maximum achievable overlap of an initial
state with a target state as a measure of a network's suitability for
certain information transmission and state preparation tasks.

First, let $\{\ket{\phi_{\In}}\}$ be an orthonormal set of input states
 (we will typically take these to be determined
by the subspace structure induced by the CSOs), which
we can use to decompose the input state
\begin{align*}
 \ket{\psi_{\text{in}}} &=\sum_n \alpha_n \ket{\phi_{\text{in}}^{(n)}}.
\end{align*}
Now suppose we can implement a particular unitary operator $U$ that maps
this set to a set of output states
$U\ket{\phi^{(n)}_{\In}}=\ket{\phi_{\text{out}}^{(n)}}$.  Expressing the
target state in this basis, we have
\begin{align*}
 \ket{\psi_{\text{targ}}}&=\sum_n \beta_n \ket{\phi_{\text{out}}^{(n)}}
\end{align*}
whereas $U$ can only transform $\ket{\psi_{\In}}$ to 
\begin{equation*}
  \ket{\psi_{\out}^U} = \sum_n \alpha_n \ket{\phi_{\text{out}}^{(n)}},
\end{equation*}
giving a maximum overlap with the target state of
\begin{equation}
\label{eqn:maxfid} \textstyle
  F = |\ip{\psi_{\out}^U}{\psi_{\targ}}|^2
    = \left|\sum_{n,m}\alpha_{m}^*\beta_{n}\right|^2.
\end{equation}
If $K=\exp(L)$ is the dynamical Lie group of the system, $L$ being
its dynamical Lie algebra, then the maximum transfer fidelity is
\begin{equation}
 \label{eqn:maxfid2} \textstyle
  F = \max_{U\in K} |\ip{\psi_{\out}^U}{\psi_{\targ}}|^2.
\end{equation}
The latter expression is not easy to evaluate in general, but we can
give upper bounds based on our knowledge of the symmetries of the
system.

For example, suppose we know that the system decomposes due to the
presence of CSOs, i.e., $\H= \bigoplus_d\H_d$ with $\dim\H_d =
N_d$ and $\sum_d N_d=N$, and the dynamics on each subspace $\H_d$ are
independent.  We have
\begin{align*}
 \ket{\psi_{\In}} 
  &= \sum_d \Pi_d\ket{\psi_{\In}} 
   = \sum_d \alpha_d \ket{\phi_{\In}^{(d)}} \\
 \ket{\psi_{\targ}} 
  &= \sum_d \Pi_d\ket{\psi_{\targ}}
   = \sum_d \beta_d \ket{\phi_{\out}^{(d)}} 
\end{align*}
where $\alpha_d=\norm{\Pi_d\psi_{\In}}$,
$\beta_d=\norm{\Pi_d\psi_{\targ}}$ and
$\ket{\phi_{\In}^{(d)}}=\alpha_d^{-1}\Pi_d\ket{\psi_{\In}}$,
$\ket{\phi_{\out}^{(d)}}=\beta_d^{-1}\Pi_d\ket{\psi_{\targ}}$ are the
respective (normalized) projections onto the subspaces.  As any vector
in the subspace $\H_d$ must remain in this subspace and the norm is
preserved under unitary evolution, the best we can hope for is to
simultaneously map the normalized states $\ket{\phi_{\In}^{(d)}}$ onto
$\ket{\phi_{\out}^{(d)}}$ for all $d$, in which case we obtain
\begin{equation}
  \label{eqn:maxfid3} \textstyle
  F=\left|\sum_d \alpha_d^* \beta_d \right|^2,
\end{equation}
which is equivalent to (\ref{eqn:maxfid}).  This bound is guaranteed to
be attainable only if all the relative phases between subspaces can be controlled, i.e., if the dynamical Lie algebras on the subspaces
are $\uu(N_d)$ or $\sp(N_d/2)\oplus \uu(1)$.  In practice, if the
subspace Hamiltonians have zero trace then we do not have (subspace)
phase controllability, and the most we can actually hope for is to map
each $\ket{\phi_{\In}^{(d)}}$ to a state
$e^{i\phi_d}\ket{\phi_{\out}^{(d)}}$, in which case we obtain the
modified bound $F=\left|\sum_d \alpha_d^* \beta_d e^{i\phi_d}\right|^2$
and the original bound may be unattainable.  Similarly, this may be the
case if the subspace dynamics is constrained by further ASOs.
Nonetheless the bound is often attainable for certain initial and target
states even in spite of these symmetries and phase constraints.

\section{Application to Spin Networks} 
\label{sec:3}

To demonstrate some of these concepts, we take the well-known class of
examples as defined by the $XX$-coupled $N$-spin networks.

Define a graph $G_{0}=(V_{0},E_{0})$ of the spin network, where $V_{0}$ is the
vertex set (of $N$ elements) of physical qubits and the edge set
$E_{0}=\left\{(i,j)|\text{qubit }i \text{ interacts directly with qubit
}j\right\}$. The system Hamiltonian for the spin network is thus 
\begin{equation}
  \label{eqn:spin_ham}
   H_0 := \frac{1}{2}\sum_{\{i,j\}\in E_{0}} d_{ij}^{(0)}\left(X_i X_j + Y_i Y_j\right), 
\end{equation}
where $X$, $Y$ and $Z$ are the standard Pauli matrices and $d_{ij}$
are real-valued coupling constants. The Hamiltonians are therefore given by real-symmetric matrices.

We can similarly add control to the network by modifying a subset of the couplings defined by the edge set of a second graph, $G_{C} = (V_{0}, E_{C})$ 
\begin{equation}
   H_C := \frac{1}{2}\sum_{\{i,j\}\in E_{C}} d_{ij}^C\left(X_i X_j + Y_i Y_j\right), 
\end{equation}
The controlled coupling can be manipulated either by a simple on-off sequence or
more complicated pulses.  It can easily be shown that the system
Hamiltonian commutes with the total spin operator:
$\left[H_{0},S_{Z}\right]=0$ for $S_Z :=\sum_{i=1}^{N} Z_{i}$ and thus
the number of excitations in the system is preserved, i.e., the dynamic
Lie algebra is a direct sum of $N$ subalgebras, each corresponding to
the evolution of $k\in\{1,\ldots, N\}$ excitations along the chain.  The
control $H_C$ also preserves the coupling isotropy in the $xy$-plane
and thus also commutes with the total spin operator, $\left[H_{C},S_{Z}\right]=0$, preserving the decomposition into excitation subspaces.  The system is therefore not controllable in a
conventional sense and would be unsuitable for $N$-qubit quantum computation
without the addition of some control that does not commute with $S_Z$ in order to
break this symmetry and couple the sub-blocks of the Hamiltonian.
However, for information transfer and state manipulations, excitation
preservation is a positive property to exploit, automatically confining
the dispersion of the excitation(s) within a lower-dimensional subspace.
This will allow us to show how a single excitation can be forced to
propagate to a chosen target state of the network. 

Within the first excitation subspace, we use the basis $\ket{n}$ to
denote an excitation localized on the $n^{th}$ qubit. Similarly, in the
second excitation subspace we use $\ket{n,m}$ to denote a pair of
excitations on qubits $n$ and $m$. In a slight abuse of this notation,
given a state $\ket{\psi}=\sum_{n=2}^N\alpha_n\ket{n}$, we will write
$\ket{1,\psi}$ to denote $\sum_{n=2}^N\alpha_n\ket{1,n}$.

We shall use the matrices $A$ and $C$ to denote the restriction of $H_0$
and $H_C$ to the single excitation subspace. $A$ is an $N\times N$ matrix
with elements $d_{ij}^{(0)}\neq 0$ for any pair $(i,j)$ which constitutes an
edge of the underlying graph (and similarly for $C$ using $d_{ij}^C$
instead). For simplicity, we shall take $d_{ij}^{(0)}>0$. If all
$d_{ij}^{(0)}=1$, then $A$ corresponds to the adjacency matrix of the
graph $G_0$. Note that all diagonal elements of both matrices are zero.

Although the following analysis could be conducted for more general
controlling interactions, in the spirit of analyzing the capabilities of
systems with minimalistic control, we consider controlling interactions
of \textit{pendant} type, i.e.~in graph $G_0$, vertex 1 is unconnected
(meaning $A_{1j}=A_{j1}=0$ for all $j$), whereas the only edge in $G_C$
is between vertices 1 and 2. Without loss of generality, we can take
$d_{12}^C=1$, corresponding to a control Hamiltonian
\begin{equation}
  \label{eqn:ctrl_ham}
 H_C := \frac{1}{2}\left(X_1 X_2 +Y_1 Y_2\right).
\end{equation}
In the first instance, we assume that the coupling strengths are all known.

We first prepare the network in the zero-excitation state using a
suitably strong background magnetic field~\footnote{Initialization using
a local-interaction cooling method~\cite{burgarthcool} can remove all
excitations except for the dark states, and we can then catalytically
reintroduce excitations to remove those dark states.}, and initialize a
state transfer or a single excitation state preparation task by
introducing a single excitation on the pendant qubit. Given the initial
state $\ket{1}$ on the pendant qubit and a target state $\ket{\psi}$ in
the single excitation subspace, what is the optimal fidelity that can be
achieved through modulation of the control $C$?

We start by applying the symmetry conditions to the particular case of a
pendant controlled spin network, restricted to the first excitation
subspace.  Let $M$ be a symmetry operator of the system.  Then
\begin{align*}
  MA \pm AM=0,\\ MC \pm CM=0,
\end{align*}
where the sign determines whether the symmetry is of ASO or CSO type
and must take the same value for both equations. 

If $M$ is a CSO, then $\ket{1}$ must be an eigenstate of $M$. Also, the
eigenvectors of $C$, $\ket{1}\pm\ket{2}$ must be eigenstates of
$M$. Combining these statements, $\ket{2}$ is an eigenstate of $M$, and
this is in the same eigenspace as $\ket{1}$. Any eigenstate of $A$ that
is in a different eigenspace of $M$ must satisfy
$\bra{1}C\ket{\lambda}=0$, and cannot be accessed. This allows us to
show that degeneracies of the system are necessarily associated with
symmetries.  If $A$ has degenerate eigenvalues then we have an
eigendecomposition $A=\sum_k\lambda_k\Pi_k$, where $\Pi_k$ are
projectors onto the respective eigenspaces, and we have either
$\Pi_k\ket{2}=0$, in which case there is no overlap of state
$\ket{2}=C\ket{1}$ with the $k$th eigenspace, or we can define an
eigenbasis such that $\ket{\lambda_{k,1}}=\Pi_k\ket{2}$.  As the
projection of a vector onto a subspace cannot have more than rank $1$,
all other eigenvectors $\ket{\lambda_{k,\ell}}$ in the $k$th eigenspace
must have zero overlap with state $\ket{2}$, and thus each
$\proj{\lambda_{k,\ell}}$ for $l>1$ commutes with both $A$ and $C$ and
is therefore a CSO. This means that when we work under the assumption of
having found all CSOs, we can take the subspaces to be
non-degenerate. Henceforth, we shall use $\mathcal{H}_a$ to denote the smallest subspace with support on $\ket{2}$, i.e.~the space that remains relevant for tasks state manipulation tasks where we start in the state $\ket{1}$ when we remove all the CSOs from the single excitation subspace. Now we wish to identify the ASOs on $\mathcal{H}_a$, as these are the only ones that could affect our state manipulation task, potentially reducing the achievable fidelity below that of Eqn.~(\ref{eqn:maxfid2}).

\begin{thm}
In a pendant-controlled system with the pendant qubit indexed as qubit 1, if $G_0$ is a graph which is connected on vertices $2$ to $N$ with positive weights, $d_{ij}^{(0)}>0$, a necessary and sufficient condition for
the existence of an ASO that applies to $\mathcal{H}_a$ is that $G_0$ is bipartite
(two-colorable).
\end{thm}
\begin{proof}
We start by noting that if $G$ is bipartite, we can easily give an
ASO within the single excitation subspace. For a graph to be bipartite, it means that we can partition the
vertices into two sets $\{V_A,V_B\}$ such that edges only connect
between the sets, not within a set. If we apply a product of $Z$
operations on one side of that bipartition, then every $(XX+YY)$
coupling is affected by exactly one $Z$ operator, with which it
anticommutes. Hence, this is an ASO, and applies in all excitation
subspaces.

In case there are additional CSOs that apply within the single excitation subspace, we need to justify that this ASO will also apply to just $\mathcal{H}_a$. Let $M$ be the matrix representing the ASO in the single excitation subspace. We can write it as the diagonal matrix
$$
M=\sum_{n\in V_A}\proj{n}-\sum_{n\in V_B}\proj{n}.
$$
As with all ASOs, $M$ imposes that the eigenvectors of $A$, $\ket{\lambda_n}$ either have 0 eigenvalue or arise in $\pm\lambda_n$ pairs with the eigenvector of eigenvalue $-\lambda_n$ being given by $M\ket{\lambda_n}$. This is proven by using the anticommutation,
$$
A(M\ket{\lambda_n}=-MA\ket{\lambda_n}=-\lambda_n(M\ket{\lambda_n}).
$$
Now let's imagine there's another CSO that applies to both $A$ and $C$. Thus, there must be a unitary matrix that acts on $A$ and $C$ to create the subspace structure. This additional structure is composed of eigenvectors $\ket{\lambda_n}$ of $A$ with $\braket{2}{\lambda_n}=0$. If these eigenvectors are subject to the ASO $M$, then the remaining subspace must also be subject to it. If $\braket{2}{\lambda_n}=0$, then is also holds that $\bra{2}M\ket{\lambda_n}=0$, so for any $\lambda_n\neq 1$, eigenvectors can be removed in pairs. These pairs manifest within $A$ or $C$ as a subspace
$$
S=\proj{\lambda_n}-M\proj{\lambda_n}M^\dagger.
$$
Since $M^2=\identity$, we have that
$$
\{S,M\}=0.
$$
On the other hand, if $\lambda_n=0$, then we just have $S=0$, and thus $\{S,M\}=0$ So, for each subspace that we remove, the matrix $M$ splits into two components, an ASO for the removed subspace, and an ASO for the remaining one.

If $M$ were to remain diagonal under the action of that unitary, then $M$ would also have the same subspace structure, and we could just extract that $\mathcal{H}_a$ component. If we are removing an eigenvector $\ket{\lambda_n}$ from the space, then it is because it has $\braket{2}{\lambda_n}=0$, which will also be true for the eigenvector $M\ket{\lambda_n}$ i.e.~for non-zero eigenvalues, the eigenvectors must be removed in pairs.

Moreover, it is the only relevant ASO in the single excitation
subspace.

\begin{lemma}
Having taken into account all CSOs, $\H_{a}$ for a pendant-controlled system has no more than one ASO.
\end{lemma}

\begin{proof}
We can take the restriction of $C$ to $\H_a$ to be
$C=\ket{1}\bra{2}+\ket{2}\bra{1}$, and the restriction of $A$ to $\H_a$ is $A'$, which has $N_a$ eigenvectors $\ket{\lambda_n}$. We can always take that $\ket{\lambda_1}=\ket{1}$, which is a 0 value eigenvector of $A'$.

Assume there exists an $M$ which anticommutes with both $C$ and $A'$.
Anticommutation with $C$ means the structure of $M$ must be such that
$\ket{1}$ and $\ket{2}$ are eigenvectors with eigenvalues of equal
magnitude but opposite sign. We can fix the scale factor of the matrix
by specifying that $\ket{2}$ is the $+1$ eigenvalue.  Anticommutation
with $A'$ implies that eigenvectors come in $\pm\lambda_n$ pairs, with a
single 0 eigenvector if $N_a$ is even.  We order the eigenvectors such
that the eigenvalues $\lambda_n=-\lambda_{n+n_a}$ for $n=2$ to $1+n_a$ with
$n_a=\lfloor(N_a-1)/2\rfloor$. Hence, we can write
$M$ as
\begin{align}
  M=-\proj{1}+\sum_{n=2}^{1+n_a} 
  \beta_n(\ket{\lambda_n}\bra{-\lambda_n}+\ket{-\lambda_n}\bra{\lambda_n}).
\end{align} 
If $N_a$ is even then $A'$ has a second $0$-eigenvalue
($\ket{\lambda_{N_a}}$) which adds a term 
$\beta_{N_a}\proj{\lambda_{N_a}}$ to $M$.  Let
$U=\sum_{n=1}^{N_a}\ket{n}\bra{\lambda_n}$ be the unitary
operator that diagonalizes $A'$.  Then $UA'U^\dag=$
\begin{align*}
  \left\{\begin{array}{ll}
  \text{diag}(0,\lambda_2,\lambda_3\ldots\lambda_{1+n_a},-\lambda_2,-\lambda_3\ldots) & N_a \text{ odd} \\
  \text{diag}(0,\lambda_2,\lambda_3\ldots\lambda_{1+n_a},-\lambda_2,-\lambda_3\ldots,0) & N_a \text{ even}.
\end{array}\right.
\end{align*}
The action of $U$ also simplifies $M$, $UMU^\dag=$
\begin{align*} 
 \left\{\begin{array}{ll}
(-1)\oplus \left(X \otimes
	    \text{diag}(\beta_2,\beta_3\ldots\beta_{1+n_a})\right) 
& N_a \text{ odd} \\
(-1)\oplus \left(X \otimes
	    \text{diag}(\beta_2,\beta_3\ldots \beta_{1+n_a})\right)
            \oplus\beta_{N_a} 
& N_a \text{ even}. 
\end{array}\right.
\end{align*}
As $\ket{2}$ is a $+1$ eigenvector, we have $M\ket{2}=\ket{2}$.
Defining 
\begin{equation*}
\ket{v}=\left\{\begin{array}{ll}
\left(1\oplus (H\otimes \identity)\right)U^\dagger\ket{2}	
& N_a \text{ odd} \\
\left(1\oplus (H\otimes \identity)\oplus 1\right)U^\dagger\ket{2}	
& N_a \text{ even}
\end{array}\right.
\end{equation*}
where $H$ is the Hadamard gate, it follows that $\ket{v}$ is a $+1$
eigenvector of $(H\otimes \identity)M(H\otimes\identity)$, which is a
diagonal matrix. Evidently, one of $\braket{n}{v}$ and
$\braket{n+n_a}{v}$ must be $0$ for $n\geq 2$.  However, both cannot be
$0$ because that would imply $\braket{n}{v}\pm\braket{n+n_a}{v}=0$, and
thus $\bra{n}U^\dag\ket{2}=\braket{\lambda_n}{2}=0$, which is not
allowed since there are no CSOs on $\H_a$.  This uniquely determines
$M$.  If $N_a$ is even, the remaining coefficient $\beta_{N_a}=1$ as we
must have $\braket{N_a}{v}=\braket{\lambda_{N_a}}{2}\neq 0$.
\end{proof}

It remains to prove that non-bipartite graphs do not have an ASO on $\mathcal{H}_a$. Let us
assume the contrary -- there exists an ASO, $M$. This means that all the
eigenvectors $\ket{\lambda_n}$ with eigenvalue $\lambda_n\neq0$ have a
counterpart $M\ket{\lambda_n}=\ket{-\lambda_n}$ of eigenvalue $-\lambda_n$. Moreover, as argued above, $\ket{2}$ is an eigenvector of the symmetry operator, so these eigenvectors have the same value of
$\alpha_n:=\braket{2}{\lambda_n}$. Following a similar route to
\cite{akreview}, this assumption imposes that
\begin{equation}
 \bra{2}A^{2k+1}\ket{2}=\sum_n|\alpha_n|^2\lambda_n^{2k+1}=0.	
 \label{eqn:odd_power}
\end{equation}
We can also expand the $\bra{2}A^{2k+1}\ket{2}$ in terms of the coupling
strengths $d_{ij}^{(0)}$; it is a sum of paths of $2k+1$ steps starting and ending on vertex 2, where for each path we just take the product of the coupling strengths along that path. The existence of such a path implies the existence of an odd cycle in the graph i.e.~for bipartite graphs, $\bra{2}A^{2k+1}\ket{2}=0$. For a non-bipartite graph, however, there do exist odd loops. If all the $d_{ij}^{(0)}>0$, then the product around any given path is positive, and the sum over all paths must also be positive i.e.~there must exist a $k$ for which $\bra{2}A^{2k+1}\ket{2}>0$, and we must see such a path if the graph is connected and all the coupling strengths are positive; Eqn.~(\ref{eqn:odd_power}) must be
violated, and there cannot be an ASO\footnote{If some coupling strengths are allowed to be negative, one can find counter-examples in which the weight on one loop can exactly cancel the weight on a second loop, such that the non-bipartite graph does have an ASO. These, however, are highly specialized cases.}. Whether we operate in the single excitation subspace or $\mathcal{H}_a$ does not matter to this argument as the only difference is the presence or absence of eigenvectors with $\alpha_n=0$, which therefore do not affect $\bra{2}A^{2k+1}\ket{2}$.
\end{proof}

\begin{thm}[Maximum transfer fidelity] \label{thm:main}
Let $A$ correspond to the single excitation subspace of an $XX$ Hamiltonian with positive weights, and a coupling geometry specified by a graph $G_0$ that is connected and not bipartite. In conjunction with a pendant-type control $C$, the maximum
fidelity for transfer of an initial state $\ket{1}$ to a target state
$\ket{\psi}$ in the single-excitation subspace is
\begin{equation}
 \label{eqn:maxfid4}
  F_{1 \mapsto \psi} 
  = 1-\sum_{\{(k,\ell)|\bra{1}C\ket{\lambda_{k,\ell}} = 0\}}
   |\ip{\psi}{\lambda_{k,\ell}}|^2,
\end{equation}
and this bound is attainable, provided $\{\ket{\lambda_{k,\ell}}\}$ is an
eigenbasis of $A$ with $\ell=1,\ldots,\text{rank}(\Pi_k)$ satisfying 
\begin{subequations}
   \begin{align}
   \label{eqn:cond1}
   \bra{1}C\ket{\lambda_{k,\ell}}=0, &\qquad \forall k,\; \forall \ell>1,
    \text{ and} \\
   \Pi_m\ket{\lambda_{k,\ell}}=0,    &\qquad \forall m\neq k&
\end{align}
\end{subequations}
where $\Pi_k$ is the projector onto the $k$-th eigenspace.
\end{thm}
\begin{proof}
Let us assume that $A$ has an eigendecomposition 
$$
  A=\sum_n \lambda_n \ket{\lambda_n}\bra{\lambda_n}.
$$
We can always take $\ket{\lambda_1}=\ket{1}$ and $\lambda_1=0$ as
$\ket{1}$ is an eigenstate of $A$ with eigenvalue $0$.  Moreover,
by the definition of the pendant-type control system the only connection
of this subsystem to the rest of the network is through state $\ket{2}$
via $C$.  Define the $C$-overlaps
\begin{equation}
  \alpha_n :=\bra{1}C\ket{\lambda_n} = \ip{2}{\lambda_n},
\end{equation} 
and let $I_0=\{n|\alpha_n \neq 0\}$.  If $\alpha_n=0$ then the subspace
spanned by $\ket{\lambda_n}$ is decoupled and cannot be accessed.
Hence, any state produced from the initial state using the control must
be a superposition of eigenstates $\ket{\lambda_n}$ that have non-zero
overlap with $\ket{2}$,
\begin{equation*}
  \ket{\phi} = \sum_{n\in I_0} \gamma_{n} \ket{\lambda_{n}}.
\end{equation*}
Writing the target state $\ket{\psi}$ in the $A$-eigenbasis $\ket{\psi}
= \sum_{n} \beta_n^{(\psi)} \ket{\lambda_n}$, the overlap of the output
state with $\ket{\psi}$ is
\begin{equation*}
   F_{1\mapsto\psi} = |\ip{\phi}{\psi}|^2 = 
    \left|\sum_{n\in I_0} \gamma_{n}^{*}\beta_n^{(\psi)} \right|^2.
\end{equation*}
As we must have $\sum_n |\gamma_n|^2=1$ and $\gamma_n=0$ for $n\not\in
I_0$, the overlap is maximised if we choose
\begin{equation*}
 \gamma_n = 
 \frac{\beta_n^{(\psi)}}{\sqrt{\sum_{n\in I_0}|\beta_{m}^{(\psi)}|^{2}}}, 
 \quad \forall n\in I_0.
\end{equation*}
in which case we obtain
\begin{align*}
F_{1\mapsto\psi} 
&= \left|\frac{\sum_{n\in I_0} |\beta_n^{(\psi)}|^2}{
   \sqrt{\sum_{n\in I_0} |\beta_n^{(\psi)}|^2}}\right|^2 
   = \sum_{n\in I_0} |\beta_n^{(\psi)}|^2\\
&= 1 - \sum_{n\not\in I_0} |\beta_n^{(\psi)}|^2 
 = 1 - \sum_{n\not\in I_0}|\ip{\psi}{\lambda_n}|^2.
\end{align*}
A non-bipartite graph has no ASOs, so we have frequency-selective
control of each transition $\ket{1}\mapsto\ket{\lambda_n}$ for which the
transition probability $|\bra{\lambda_1}C\ket{\lambda_n}|^2$ does not
vanish by using a Rabi oscillation of the control field $C$ with
frequency $\omega_{1n}=\lambda_n-\lambda_1=\lambda_n$. There are some
special cases which we have to handle. It can be the case that there are
two eigenvectors with eigenvalue $\pm\lambda$. Provided that the
$|\alpha_n|$ are distinct, a Rabi oscillation still suffices via
judicious choice of pulse timing. If the $|\alpha_n|$ are also equal,
one has to use a different technique -- we find a different level
$\ket{\kappa}$ which we populate first, and then perform a two frequency
(Raman) transition with frequencies $\lambda-\Delta$ and
$\kappa-\Delta$, where $\Delta$ is a detuning. The only way that this
technique can fail is if the system is so structured that all
eigenvectors satisfy this pairing property, or have 0 eigenvalue
i.e.~there is an ASO present, but we know there is no such ASO.

 We can therefore, in principle, transform the state $\ket{1}$ to the output
state $\ket{\phi}$ by applying a sequence of pulses resonant with the
transition frequencies $\lambda_n$ as in~\cite{ak_and_me}, for instance. With arbitrarily slow and weak pulses, the off-resonant excitations are arbitrarily weak, however off-resonant excitations could be suppressed, transfer times reduced, and other constraints introduced, by using optimal control pulse design~\cite{ramil,rap_comm_2009}.
In the case of an on-off control switch, we could try to match the
frequencies in the Fourier decomposition of the square pulses with the
resonant frequencies as pointed out in \cite{ak_and_me}.
\end{proof}

In the case of bipartite graphs, there is an ASO present, and attempting a Rabi oscillation using
$C$ to address the transition between $\ket{1}$ and $\ket{\lambda_n}$ equally couples to the transition between $\ket{1}$ and
$\ket{-\lambda_n}$, so we can only prepare paired eigenvectors in states
$$
e^{i\lambda_nt}\ket{\lambda_n}+e^{-i\lambda_nt}\ket{-\lambda_n}
$$
i.e.~they have the same mod-square amplitude. In a similar way to Example 1, we will prove that this happens however we attempt to control the system, and give a physical interpretation of the states satisfying this property.
\begin{lemma}
A state transformation task on a bipartite pendant controlled graph can be implemented up to the bounds specified in Theorem \ref{thm:main} if and only if the target state has (up to a global phase) real amplitudes on one of the bipartitions, $V_A$ and imaginary amplitudes on the other, $V_B$.
\end{lemma}
\begin{proof}
The states that we can make are of the form
$$
\ket{\psi_{\text{targ}}}=\sum_n\beta_n(e^{i\phi_n}\ket{\lambda_n}+e^{-i\phi_n}\ket{-\lambda_n})
$$
with real $\beta_n$. The bipartite graph in the subspace $\mathcal{H}_a$ (any weight of the target state outside this space is explicitly forbidden by Theorem \ref{thm:main}) has exactly one ASO. It suffices for us to describe its effect on the single excitation subspace since we have already seen how this symmetry splits under the action of any further CSOs.
$$
M=\sum_{n\in V_A}\proj{n}-\sum_{n\in V_B}\proj{n}.
$$
Since all eigenvectors $\ket{\lambda_n}$ of $A$ have $M\ket{\lambda_n}=\ket{-\lambda_n}$, $\ket{\lambda_n}$ and $\ket{-\lambda_n}$ have real amplitudes $\gamma_{nm}$ with equal modulus on every spin $m$, and the relative phase in amplitude, $\pm1$, is determined entirely by which side of the bipartition they sit on. Hence,
\begin{eqnarray}
\ket{\psi_{\text{targ}}}&=&2\sum_{m\in V_A}\sum_n\beta_n\gamma_{nm}\cos(\phi_n)\ket{m}	\nonumber\\
&&+2i\sum_{m\in V_B}\sum_n\beta_n\gamma_{nm}\sin(\phi_n)\ket{m}.	\nonumber
\end{eqnarray}
All the amplitudes on $V_A$ are real, all those on $V_B$ are imaginary and all such amplitudes are reachable.

To see that the transformation is otherwise impossible, consider the operator
$$
U_B =\prod_{n\in V_B}\sqrt{Z_n}.
$$
If we transform $H_0$ and $H_C$ according to $U_B$, all $(XX+YY)$ terms are transformed into $(XY-YX)$ terms, which are imaginary. Hence, all evolutions $-iU_B (H_0+f(t)H_C)U_{B}^\dagger$ are real, and the corresponding unitary operators are real. By starting with a state $U_B \ket{1}=\ket{1}$, which is real, the amplitudes of all output states $U_B \ket{\psi}$ must be real. Thus $\ket{\psi}$ must be real on spins that are in $V_A$, and imaginary on those in $V_B$.
\end{proof}

In particular, this means that state transfer tasks are unaffected by ASOs. The phase result makes a lot of sense because we know that evolution under a Hamiltonian $\half(XX+YY)$ leads to a hopping term between two vertices, but introduces a phase factor of $-i$ for each hop. Hence, on a bipartite graph, if we start on a single site with a real amplitude, all those that are an even number of hops away (which is only well defined on a bipartite graph) must have a real amplitude, whereas those that are an odd number of hops away must have an imaginary amplitude.

Alternatively, the limitation is easily overcome -- we can introduce a second control
field $C'=i\ket{1}\bra{2}-i\ket{2}\bra{1}$, which corresponds to a
coupling $\half(X_1Y_2-Y_1X_2)$.  As this coupling is not real, it does not satisfy (\ref{eqn:antisymmfull}), although it does still anticommute with the symmetry operator. Using a Rabi oscillation of
$\cos(\omega_{1n}t)C\pm\sin(\omega_{1n}t)C'$ allows us to selectively
make the transition of one or other of the eigenvectors. This illustrates the important difference between the symmetry condition of Eqn.~(\ref{eqn:antisymmfull}) and anticommutation if the Hamiltonian is not real-valued. With this additional control in place, the conditions of Thm.~\ref{thm:main} are all achievable. However, we will not further consider this control.

In summary, we see that in all cases the maximum possible state transfer
fidelity is entirely determined by the energy eigenstates of $A$ which
have zero overlap with $\ket{2}$, the point where the pendant control
meets the rest of the system. Entanglement generation, however, is additionally affected in bipartite graphs by whether the pair of qubits that are to be entangled are separated by an even or odd number of edges, and the relative phase to be generated between the pair -- we can either make $\ket{n}+\ket{m}$ for even distances, or $\ket{n}+i\ket{m}$ for odd distances.

\begin{figure}
 \centering
 \includegraphics[width=208pt,height=72pt]{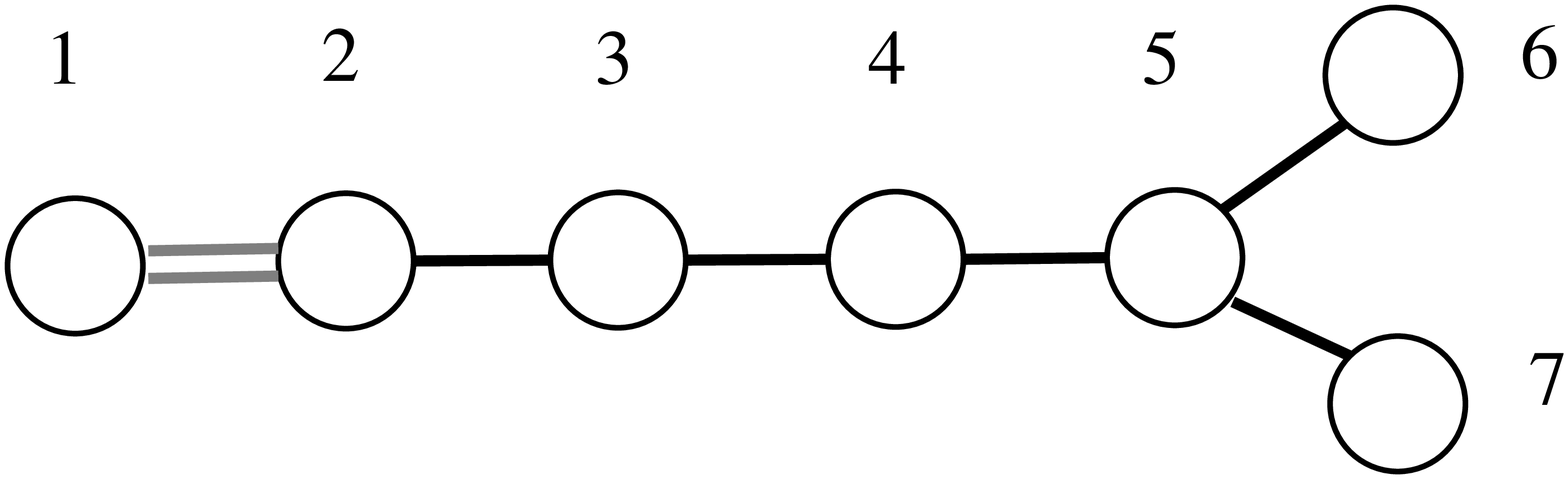}
 \caption{Controlled interaction between spins 1 and 2 marked double in gray}
 \label{Fig1}
\end{figure}

\noindent
\textbf{Example 2:} First we examine a uniformly $XX$-coupled $N$-spin
chain with a fork at the end (see Fig.~\ref{Fig1}), initialized with a
single excitation localized on the first qubit.  Diagonalizing the
adjacency matrix shows that the only eigenstate with zero $C$-overlap is
$\ket{\lambda_{-}} = \frac{1}{\sqrt{2}}(\ket{6}-\ket{7})$.  Consider
state transfer to $\ket{6}$; when we write this in the adjacency matrix
eigenbasis, the coefficient of the $\ket{\lambda_{-}}$ eigenstate is
$\beta_{-}^{(6)} = 1/\sqrt{2}$, giving a maximum transfer fidelity $F_{1
\mapsto 6} = 1-|\beta_{-}^{(6)}|^{2}=1-|\frac{1}{\sqrt{2}}|^{2} =
\frac{1}{2}$; the same result is obtained for transfer to $\ket{7}$ by
symmetry. This is to be expected; the two end qubits are in some sense
indistinguishable, and we cannot direct information flow to either one
specifically. At all times, the control addresses both simultaneously,
and so we obtain a superposition state over these two qubits.  This is
made clear in the algorithmic computations. In this case, the Hamiltonian
is symmetric under the change of basis CSO $J_{67}$ taking $J_{67}\ket{6}=\ket{7}$,
$J_{67}\ket{7}=\ket{6}$, $J_{67}\ket{n} = \ket{n}$ otherwise.
Performing the computations above shows that the Hamiltonian in this
subspace generates a dynamical Lie algebra of $\so(6)\oplus\identity$,
where the $\identity$-term corresponds to the subspace spanned by the
dark state $\frac{1}{\sqrt{2}}(\ket{6}-\ket{7})$.  If our initial state
has no overlap with this state then it remains unpopulated and ``dark''
throughout.  Alternatively if our initial state is contained in this
subspace, it will be preserved throughout the system's evolution. On the other hand,
we can choose to produce states such as $(\ket{6}+\ket{7})/\sqrt{2}$,
which is an entangled state, or states such as $\ket{5}$, which would
allow perfect transfer of an unknown qubit state from qubit 1 to qubit
5. However, without the additional control $C'$, it is impossible to perfectly produce states such as $\frac{1}{\sqrt{2}}(\ket{3}+\ket{4})$ as a result of the bipartite nature of the graph, and the disruption caused by the ASO.

\begin{figure}
 \centering
 \includegraphics[width=208pt,height=72pt]{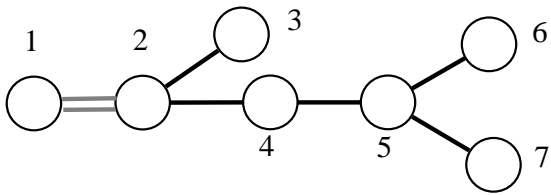}
 \caption{Controlled interaction between spins 1 and 2 marked double in gray}
 \label{Fig2}
\end{figure}

The results in the previous example are intuitive, but, for other
systems, identifying the symmetries and dark subspaces may be less
intuitive.

\noindent\textbf{Example 3:} If we make a small modification to the
system in Fig.~\ref{Fig1} by moving the pendant vertex as shown in 
Fig.~\ref{Fig2}, then additional symmetries and dark states arise.  
Since the graph is bipartite, with partitions $V_1=\{1,3,4,6,7\}$ and $V_2=\{2,5\}$, we can immediately observe that there is an ASO
$$
M=\identity-2(\proj{2}+\proj{5}),
$$
derived from $Z_2Z_5$,
which anticommutes with the Hamiltonian. We again observe the permutation CSO
$$
J_{67}=\identity-(\ket{6}-\ket{7})(\bra{6}-\bra{7}),
$$
which commutes with the Hamiltonian. These two symmetry operators give rise to an eigenspace of eigenvalue zero which has no overlap with $\ket{2}$ and is spanned by the vectors
\begin{align*}
  \ket{\lambda_{0,1}} 
   &=\frac{1}{\sqrt{2}}(\ket{6}-\ket{7}),\\
  \ket{\lambda_{0,2}} 
   &= \frac{1}{\sqrt{10}}(2\ket{3}-2\ket{4}+\ket{6}+\ket{7}).
\end{align*} 
No other eigenvectors have zero overlap with $\ket{2}$ thus the maximum transfer fidelity to $\ket{3}$, for example, is
$F_{1\mapsto3} = 1-|\beta_1^{(3)}|^2-|\beta_2^{(3)}|^2=\frac{3}{5}$ as
$\beta_{1}^{(3)}=0$ and $\beta_{2}^{(3)} =
\frac{2}{\sqrt{10}}$.  Similarly, the state
$\frac{1}{\sqrt{2}}(\ket{6}+\ket{7})$ can only be produced with fidelity
$1-\frac{1}{5}=\frac{4}{5}$.

\section{Catalytic excitations and the classification of dark states} 
\label{sec:dark_states}

So far we have discussed access to states in the first-excitation
subspace using paths entirely within this subspace, as assumed
universally in prior works.  The controlling interaction was
assumed to commute with the total spin operator $S_Z$ and was thus
unable to produce paths which move between excitation subspaces. To
access dark states, we could modify the control so that it does not
commute with $S_Z$, breaking the symmetry and thus the excitation
subspace structure.  Doing so could lead to difficulties: the resulting
subspaces will be of much larger dimension, and this will increase the
computational difficulty of finding controls.  The higher dimensional
systems are more likely to contain uncharacterized resonances which
makes it harder to produce controls which confine the excitations to the
desired path in state space.  However, we shall now show that moving to
higher excitation subspaces can sometimes be advantageous and need not
complicate the control too much.

In general, breaking symmetries requires changing the control
Hamiltonians or adding new controls.  However, in some cases we can
avoid certain symmetries by moving to higher excitation subspaces, using
our ability to introduce additional excitations on the first spin, which
has been implicitly present as a control within the initialization
stage, but we have not explicitly considered it when deriving accessible
states. To produce a desired target state $\ket{f_k}$ in the
$k$-excitation subspace, we may be able to bypass symmetry-induced
restrictions in this subspace by temporarily introducing additional
excitations, which we shall call \emph{catalytic} excitations.  To do
so, we pass to the $l$-excitation subspace which contains a state of the
form $\ket{f_k}\otimes\ket{\phi}$ in its reachable set, and finally
extract the excess excitations to leave $\ket{f_k}$.  Not all symmetries
can be avoided this way; permutation symmetries which do not act on
qubit 2 certainly cannot be avoided. However, in many cases, adding a
single extra excitation will suffice.

From \cite{osborne}, we have
\begin{definition}
The matrix representing the second excitation subspace of $H_0$ is given by
\begin{multline}
 \Lambda^2 (A)_{mn} :=\\ 
 \frac{1}{4}|(\bra{k,l}-\bra{l,k})
 \left[I_N \otimes A + A \otimes I_N \right](\ket{i,j}-\ket{j,i})|	
\label{eqn:twovertex}
\end{multline}
for $m:=(N-1)k+l$ and $n:=(N-1)i+j$, where $\ket{i,j}$ denotes the
vertices at which the excitations are localized.
\end{definition}

Similarly, $\Lambda^2(C)$ is the action of the control upon
this subspace.  The connectedness of this graph under basis
changes (and thus the irreducibility of the dynamical Lie algebra) is
not easily determined from the connectedness of $A$ and $C$ \footnote{If
it were not for the modulus in Eqn.~(\ref{eqn:twovertex}), we could
directly construct some of the symmetry operators in higher excitation
subspaces directly from those in lower excitation subspaces, but this
would make catalysis impossible. The only graph for which all these
terms are positive anyway is the chain.}.  In other words, the existence, or lack thereof,
of ASOs and CSOs other than permutation symmetries in one excitation
subspace does not necessarily imply their existence in another
excitation subspace, therefore we must search for them in each subspace
separately.  It is this feature that potentially enables catalysis,
illustrating that, in general, it is not sufficient for analyses of
state transfer in spin networks to restrict to the single excitation
subspace.

\begin{thm}[Sufficient condition for increased control through catalytic excitations]
A system has an accessible pair
$(\ket{\psi_{\text{in}}},\ket{\psi_{\text{out}}})$ in the single
excitation subspace using catalytic excitations (assuming
$\braket{1}{\psi_{\text{out}}}=0$) if it has accessible pairs
$(\ket{\psi_{\text{in}}},\ket{\psi'})$ and
$(\ket{1,\psi'},\ket{1,\psi_{\text{out}}})$ with
$\braket{1}{\psi'}=0$. \label{thm:catalytic}
\end{thm}

\begin{proof}
Initialize the system in the state $\ket{1}$ and transform it to the
state $\ket{\psi'}$ in the single-excitation subspace, which by
hypothesis is accessible and satisfies $\braket{1}{\psi'}=0$.
Introducing a second \textit{catalytic} excitation at the first site
then produces the state $\ket{1}\otimes\ket{\psi'}$ in the second
excitation subspace, and by hypothesis, we can create the state
$\ket{1}\otimes\ket{\psi_{\text{out}}}$.  Removing the catalytic
excitation on the first site, we are left with our desired target state,
$\ket{\psi_{\text{out}}}$.
\end{proof}

We now give an example to prove that such a situation can arise, even
when $(\ket{\psi_{\text{in}}},\ket{\psi_{\text{out}}})$ are not an
accessible pair when confined to evolution solely within the single
excitation subspace.

\noindent \textbf{Example 4:} For the system given in Fig.~\ref{Fig2},
we saw in Example 3 that the target state $\ket{3}$ is not perfectly
accessible from the initial state $\ket{1}$.  As before, we look at the
eigenbasis of $A$ where we have eigenvectors corresponding to
eigenvalues of $\lambda_0 = 0$ and $\lambda_{\pm,\pm}=\pm \sqrt{(5 \pm
\sqrt{5})/2}$.  Labelling the corresponding eigenvectors
$\ket{\lambda_0}$ and $\ket{\lambda_{++}}$, $\ket{\lambda_{-+}}$,
etc. we have
\begin{equation*}
\begin{split}
 \ket{3} = \sqrt{\frac{2}{5}}\ket{0} 
 +\frac{\sqrt{10}-\sqrt{2}}{4\sqrt{5}}(\ket{\lambda_{++}}+\ket{\lambda_{-+}})\\ 
 +\frac{\sqrt{10}+\sqrt{2}}{4\sqrt{5}}(\ket{\lambda_{+-}}+\ket{\lambda_{--}})
\end{split}
\end{equation*}
and starting from $\alpha\ket{0}+\beta\ket{1}$, the state with maximum
overlap with $\ket{3}$ we can achieve is
\begin{equation*}
 \alpha\ket{0} + 
 \beta\left[\frac{\ket{3}-\sqrt{\frac{2}{5}}(\sqrt{\frac{2}{5}}\ket{3}
 -\sqrt{\frac{2}{5}}\ket{4}
 +\sqrt{\frac{1}{10}}(\ket{6}+\ket{7}))}{\sqrt{\frac{3}{5}}}\right]
\end{equation*}
with maximum fidelity \footnote{Notice that
$\alpha\ket{0}+\beta\ket{\psi}$ is a superposition of the
zero-excitation state $\ket{0}$ and the state $\ket{\psi}$ in the single
excitation subspace, and the relative phase of the superposition state
is determined by the global phase of $\ket{\psi}$.  This is a problem
since we usually have limited control over the global phase of
$\ket{\psi}$.  This has been a perennial problem in the study of state
transfer, and it is always assumed that one can simply correct for this
phase difference with a local phase gate.}
\begin{equation*}
  \sqrt{F_{(\alpha\ket{0}+\beta\ket{1}) 
  \mapsto (\alpha\ket{0}+\beta\ket{3})}} 
 = |\alpha|^2+|\beta|^2\sqrt{\frac{3}{5}}.
\end{equation*}
Instead of this, we could first produce 
\begin{equation*}
\begin{split}
 \alpha\ket{0}+ \beta\Big[\frac{\sqrt{10}+\sqrt{2}}{4\sqrt{5}}(\ket{\lambda_{+-}}+\ket{\lambda_{--}}) \\
+\sqrt{\frac{1}{10}(7-\sqrt{5}})(\ket{\lambda_{++}}+\ket{\lambda_{-+}})\Big],
\end{split}
\end{equation*}
defining $\ket{\psi'}$, and then introduce a single \textit{catalyst
excitation} at spin 1,
\begin{equation*}
\begin{split}
 \alpha\ket{1}+ \beta\Big[\frac{\sqrt{10}+\sqrt{2}}{4\sqrt{5}}(\ket{1,\lambda_{+-}}+\ket{1,\lambda_{--}})\\
+\sqrt{\frac{1}{10}(7-\sqrt{5}})(\ket{1,\lambda_{++}}+\ket{1,\lambda_{-+}})\Big].
\end{split}
\end{equation*}

\begin{figure}
 \centering
 \includegraphics[width=220pt,height=140pt]{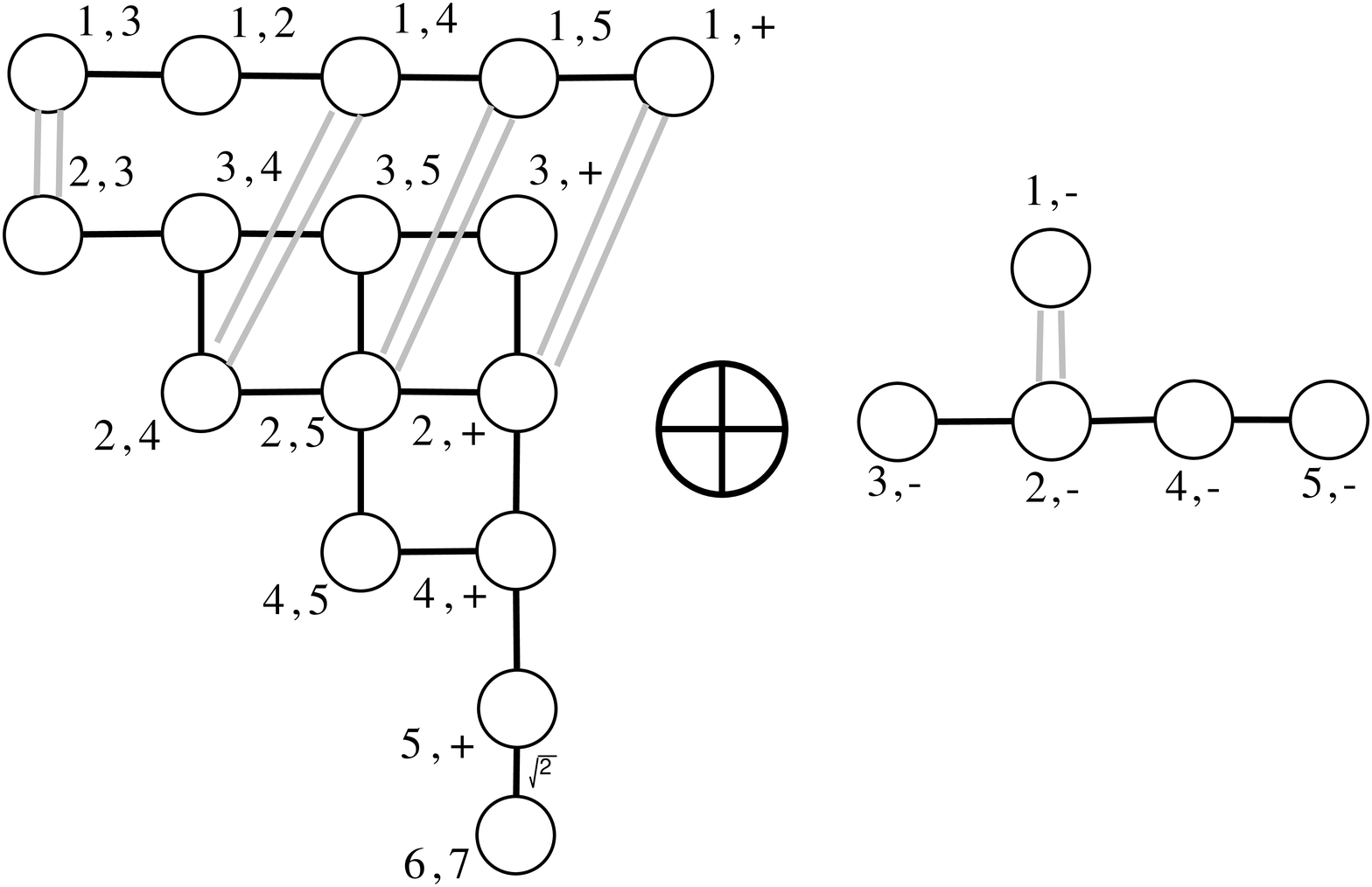}
 \caption{Controlled interactions marked double in gray; i,j represents product state $\ket{i}\ket{j}$; +~represents $\frac{1}{\sqrt{2}}(\ket{6}+\ket{7})$, -~represents $\frac{1}{\sqrt{2}}(\ket{6}-\ket{7})$. Note that the controlled couplings are simultaneously controlled, not independently.}
 \label{Fig. 3}
\end{figure}

We have moved into a superposition of the 1 and 2-excitation subspaces,
and our control will keep the state in these subspaces.  To show that we
can reach the target state using catalytic excitations we need to prove that the set of states
\begin{equation*}
 \{(\ket{1,\psi'},\ket{1,\psi_{\text{out}}}), (\ket{1},\ket{1})\}
\end{equation*}
are simultaneously accessible such that the $\ket{0}$ component of the
unknown state is also correctly transferred.  This 2-excitation
subspace has no dark states which are left uncoupled to the controlling
interaction, except those which are a product of
$\frac{1}{\sqrt{2}}(\ket{6}-\ket{7})$ and a single excitation on one of
qubits 1 to 5.  Examining the graph produced by the Hamiltonian for
the 2-excitation subspace (Fig.~\ref{Fig. 3}, formed as in
\cite{osborne}), we can see that there are no eigenstates of the
Hamiltonian for the left-hand component (i.e.~the connected component to
our initial state) that have no overlap with all the states our control
connects to.  This is the analogue to checking which states had
$\bra{1}C\ket{\lambda_n}=0$ above.

We select an eigenvector $\ket{{\tilde \lambda}}$ in the second
excitation subspace. Based on this choice, we apply a field of
$d_{12}\propto \cos((\lambda_{++}-\tilde{\lambda})t)$ on the controlled
coupling between spins 1 and 2, and, neglecting off-resonant
excitations, we obtain an effective coupling
\begin{equation*}
\begin{split}
 H_1 \propto
 &\ket{1,\lambda_{++}}\bra{\tilde{\lambda}}+\ket{\tilde{\lambda}}\bra{1,\lambda_{++}}\\ &+\ket{1,\lambda_{-+}}\bra{\tilde{-\lambda}}+\ket{\tilde{-\lambda}}\bra{1,\lambda_{-+}}
\end{split}
\end{equation*}
which can transform the state to
\begin{equation*}
\begin{split}
 \alpha\ket{1}+\beta\Big[\frac{\sqrt{10}+\sqrt{2}}{4\sqrt{5}}(\ket{1,\lambda_{+-}}+\ket{1,\lambda_{--}})+\sqrt{\frac{1}{5}}(\ket{\tilde{\lambda}}\\
+\ket{\tilde{-\lambda}})+\frac{\sqrt{10}-\sqrt{2}}{4\sqrt{5}}(\ket{1,\lambda_{++}}+\ket{1,\lambda_{-+}})\Big].
\end{split}
\end{equation*}
Similarly, neglecting off-resonant excitations, applying a field 
$J_{12} \propto \cos(\tilde{\lambda}t)$ on the controlled transition
gives rise to an effective Hamiltonian
\begin{align*}
 H_2 \propto 
 \ket{1,\lambda_{0}}\bra{\tilde{\lambda}}+\ket{\tilde{\lambda}}\bra{1,\lambda_{0}}
 +\ket{1,\lambda_{0}}\bra{-\tilde{\lambda}}+\ket{-\tilde{\lambda}}\bra{1,\lambda_{0}}
\end{align*}
which can transform the previous state into
\begin{align*}
 \alpha\ket{1}+\beta\Big[\frac{\sqrt{10}+\sqrt{2}}{4\sqrt{5}}
 (\ket{1,\lambda_{+-}}+\ket{1,\lambda_{--}})\\ 
 +\sqrt{\frac{2}{5}}\ket{1,\lambda_0}+\frac{\sqrt{10}-\sqrt{2}}{4\sqrt{5}}(\ket{1,\lambda_{++}}+\ket{1,\lambda_{-+}})\Big] \\ 
=\alpha\ket{1}+\beta\ket{1,3}.
\end{align*}
Removing the catalyst excitation from the first spin gives the desired
final state, $\alpha\ket{0}+\beta\ket{3}$.  Introducing this excitation
allowed us to move between different subspaces induced by the symmetry.

Notice that the catalyst
excitation cannot make the state $\frac{1}{\sqrt{2}}(\ket{6}-\ket{7})$
accessible from an initial state $\ket{1}$ because the graph derived
from the Hamiltonian will have two components, with one component
comprising states of the form $\ket{\psi}\otimes
\frac{1}{\sqrt{2}}(\ket{6}-\ket{7})$ in all the excitation subspaces.
Therefore, we can say that there are two types of dark state, one of
which one is not as dark as the other.  Dark states arising from a permutation of the qubits \cite{tosh} are truly
dark -- they induce a splitting of the Hamiltonian which persists in all
excitation subspaces, and which cannot be overcome without introducing a
symmetry-breaking control.  The second class of dark states are
artifacts of the particular eigenvectors found in each subgraph, whose
underlying symmetries may be bypassed in higher excitation
subspaces, potentially rendering them accessible.  However, it
should be noted that if $A$ is a simple one-dimensional chain then these
symmetries do persist, as can be proven via the Jordan-Wigner
transformation \cite{ak_and_me}.

\section{System Identification} \label{sec:5}

Our control method in Sec.~\ref{sec:3}, based on geometric control, is
sub-optimal and, if the system Hamiltonian is known, there are various
ways to derive control pulses that saturate the bounds. However, these
require some basic knowledge of the Hamiltonian. In particular, given
the promise of pendant control, we only need to know the eigenvalues
$\lambda_n$, $\alpha_n$ (the overlaps of the eigenvector with $\ket{2}$,
the single excitation on spin 2), and some information about the target
state -- its decomposition in terms of the eigenvectors.  This
decomposition must be specified with a particular phase convention for
the eigenvectors, which we shall choose such that $\alpha_n > 0$ for all
$n$.  Numerical simulations further show that it is possible to derive
pulses to achieve this threshold even when the Hamiltonian of the system
has not been determined, using optimization over the record of multiple
single-basis projective measurements on the target qubit, as developed
in~\cite{rap_comm_2009}. Indeed, we shall now prove that it is possible
to identify the control parameters $\lambda_n$ and $\alpha_n$ so that,
if given a target state expressed in terms of the eigenvector
decomposition, the state transformation task can be achieved without any
further promises on the system.

For this, consider the same drift Hamiltonian $A$ and control field $C$
that we have used so far within the single excitation subspace. By
determining the eigenvalues $\lambda_n$ and overlaps
$\alpha_n=\braket{2}{\lambda_n}$ of $A'$, the restriction of $A$ to the space $\mathcal{H}_a$, then if we are given a state
transformation task starting from $\ket{1}$, with an output state
described in terms of the amplitudes of different eigenvectors, we can
implement that transformation optimally. Of course, in order to
translate from a specification in terms of the spin basis will require
further promises about the structure of $A$. To date
\cite{burgarthgateway, new_burgarth}, such assumptions represent a
massive restriction, so it is interesting to identify what aspect of the
protocol it is that requires such assumptions.

Let us define 
\begin{align*}
  H=A+\epsilon C,
\end{align*} and assume that $H$ has eigenvectors
$\ket{\eta_n}$.  Our level of control can certainly allow us to measure,
as a function of time, quantities such as 
\begin{align*} 
 \left|\bra{1}e^{-iHt}\ket{1}\right|^2=\left|\sum_n|\braket{1}{\eta_n}|^2e^{-i\eta_n t}\right|^2, 
\end{align*} 
a Fourier transform of which will reveal the eigenvalue differences $|\eta_n-\eta_m|$  and
overlaps $|\braket{1}{\eta_n}|^{2}$. By assuming $\epsilon$ to be small, we can perform a perturbative
expansion, expressing these quantities in terms of the properties of $A'$
that we wish to determine. In particular, for $n\neq 1$ and for $\lambda_n\neq 0$, we can write
that, to first order, $\eta_n=\lambda_n$ and 
\begin{align*}
\ket{\eta_n}=
 \ket{\lambda_n}+\frac{\epsilon\bra{1}C\ket{\lambda_n}}{\lambda_n}\ket{1},
\end{align*} 
i.e., $|\braket{1}{\eta_n}|=\epsilon\alpha_n/\lambda_n$, so the $\alpha_n$ and $|\lambda_n-\lambda_m|$ are readily extracted for any eigenvector which does not have $\alpha_n=0$ i.e.~as one would expect, we cannot learn the properties of the states outside the space $\mathcal{H}_a$. If there is an eigenvector with $\lambda_m=0$ and $\alpha_m\neq 0$, then we have to apply degenerate perturbation theory, but this just means the two 0 energy levels (including $\ket{1}$) mix, with energies $\pm\epsilon\alpha_m$. Again, the relevant quantities can be extracted.

From the differences $|\lambda_n-\lambda_m|$, the eigenvalues can only be identified up to an overall global shift, $\lambda_1$, and the difference between $\lambda_n-\lambda_1$ and $-(\lambda_n-\lambda_1)$ cannot be distinguished. The global shift only alters the global phase in the subsequent state transformation protocol, and is therefore largely irrelevant, but we can still argue how to determine both of these features. Firstly, observe that neither affect a system with an ASO -- we know that $H=-H$. Hence, we only have to deal with the case where the levels can, in principle, be addressed. For a given eigenvector $\ket{\lambda_n}$, we know $\alpha_n$, but not the eigenvalue. By determining two such values, all ambiguity can be removed. If we have an eigenvector $\ket{\lambda_n}$ with eigenvalue $\lambda_n$, then provided there is not also an eigenvector with $-\lambda_n$ (if there is, we can adapt for it), we can perform a Rabi oscillation of $C$ with frequency $\lambda_n$, and in some fixed time, as specified by $|\alpha_n|^2$, we return to our initial state, having gone down to a probability of $|\alpha_n|^2$ of being found in the state $\ket{1}$ half way through the protocol. Hence, for a unique value of $|\alpha_n|^2$, we can scan through all possible frequencies and find the relevant value of $|\lambda_n|$. In order to discover the sign of this energy (since a Rabi oscillation addresses $\pm\lambda_n$ simultaneously), we instead prepare $\ket{0}+e^{i\phi}\ket{1}$ for various values of $\phi$, and transform it to $\ket{0}+e^{i\phi}\ket{\lambda_n}$, leave it for some time $t$ so that it evolves to $\ket{0}+e^{i(\phi+\lambda_nt)}\ket{\lambda_n}$, and then return it to the initial state $\ket{0}+e^{i(\phi+\lambda_nt)}\ket{1}$. By measuring the probability $p=\cos^2(\half(\phi+\lambda_nt))$ of getting the $\ket{+}$ measurement result, then provided $|\lambda_n|t\leq \pi$, the sign of 
$$
-\left.\frac{dp}{d\phi}\right|_{\phi=0}
$$
is the same as that of $\lambda_n$.

This analysis reveals that within a pendant controlled system, it is, in principle, possible to extract the values $\lambda_n$ and $\alpha_n$ that we need to know, although in practice one would undoubtedly develop more sophisticated protocols which do not rely on measuring extremely small quantities. Consequently, we observe that to implement a complete identification protocol such that we can produce a state described in the position basis rather than the eigenbasis of $A$, it is necessary to be given sufficient information on the system to be able to derive the eigenvectors from the values $\lambda_n$ and $\alpha_n$. A chain is a particularly natural candidate and can be extended to chain-like systems \cite{burgarthgateway, new_burgarth}, although one could come up with many other variants

\section{Conclusions}

In this paper, we have demonstrated how the capability of a spin network
for state transfer can be computed, which has obvious potential bearing
on the design of quantum information routers. All CSOs produce states
which are robust to dephasing noise on the pendant control spins, and
those for which catalysis fails (such as permutations) produce states
which are robust to all noise on the pendant control spins.  The
difference between the two can be interpreted in terms of selection
rules \cite{BRS}: the selection rule preventing population of some
states can be overcome by shifting to a different excitation subspace,
but superselection rules prevent access to the truly dark
states. This could direct the development of spin network memories in
terms of decoherence-free subspaces, i.e.~the dark states produced by
the permutation symmetries of the excitation subspace graph could be
used to store states.  Truly dark states will be robust to the
introduction of extra excitations in the system, but will need a
corresponding permutation symmetry-breaking local control to enable
their initialization and collection.  The weaker class of dark states
will be immediately vulnerable to the introduction of excitations
elsewhere in the system; nevertheless they will be better protected than
non-dark states and correspondingly easier to access using local
controls than the truly dark states. The suitability of each method of
storage will depend on the architecture employed and the relative
importance placed on access and permanence.  This is somewhat analogous
to the relative advantages of RAM and ROM in conventional silicon
computing.

We have seen how, in the pendant controlled spin networks, the necessary
and sufficient condition for the existence of an ASO (on $\mathcal{H}_a$) is that the
underlying graph should be bipartite. This link with the underlying
graph structure is extremely interesting because it is much less
artificial than previous studies which have found such a relation
\cite{simsev1}. These studies imposed a direct relation between the
Hamiltonian and the adjacency matrix of the graph by setting them equal,
whereas our only constraint is that the graph structure specifies where
coupling coefficients take on a non-zero value.

In the future it will be interesting to see when the state with optimal
overlap can be achieved for arbitrary Hamiltonians, not just those with
pendant controllers. A trivial extension of the present study is one where the graph $A$ is composed of two connected components, and the control $C$ couples between a single vertex of each of the two components. Further consequences of the control required to
introduce a state to transfer should also be examined \cite{peter_last}.

\section{Acknowledgements}

PJP acknowledges funding from EPSRC grant EP/D07192X/1 and thanks
T.~Schulte-Herbr\"{u}ggen and D.~Tannor for useful discussions.  ASK is
supported by Clare College, Cambridge.  SGS acknowledges funding from
EPSRC ARF Grant EP/D07192X/1 and Hitachi. This research was supported in
part by the National Science Foundation under Grant No.~PHY05-51164
(KITP).

\appendix
\section{Identification of Symmetries}
\label{appendix:A}

Given the Hamiltonian $H_0$ of an $N$-dimensional quantum system and its
$M$ control Hamiltonians $H_m$, it is relatively easy to determine the
symmetries it possesses and thus limit the possible operations that can
be realized.  First we check for decomposability into a block-diagonal
structure (i.e.~we find the CSOs).  This can be done by transforming
each of the $M$ Hamiltonians $H_m$ from its $N \times N$ matrix
representation into its Liouville representation, where it is
represented by an $N^2 \times N^2$ matrix $L^{(m)}$ with entries
\begin{equation}
   L_{s,r}^{(m)} := (H_m)_{n,k}\delta_{j,\ell} - (H_m)_{j,\ell}\delta_{n,k}
\end{equation}
for $r=(j-1)N+k$ and $s=(\ell-1)N+n$.  Repeating this for all system and
control Hamiltonians $H_m$ and stacking the resulting matrices $L^{(m)}$
vertically gives an $(M+1)N^2 \times N^2$ matrix.  Any standard routine
can be used to calculate the null space of this matrix, and the column
vectors of length $N^2$ spanning the null space, ``unstacked'' into a $N
\times N$ matrix, to give the CSOs $J$ that simultaneously
commute with all $H_m$, $\left[H_m,J \right]=0$.  This is a conceptually
very simple approach.  More efficient alternatives for simultaneous
block-diagonalization such as the algorithm in \cite{simuldecomp}, which
requires handling matrices of lower dimension, exist.

To find the ASOs, we go through much the same procedure:
for each indecomposable subspace, indexed by $d$, of each Hamiltonian $H_m$, we take the complexified system and
control Hamiltonians restricted to the subspace, $iH_{m,d}$, remove the
trace
\begin{equation}
  \tilde{H}_{m,d} := iH_{m,d} - \Tr(iH_{m,d})\frac{\identity}{N},
\end{equation}
and compute the Liouvillian representation 
\begin{equation}
  \tilde{L}_{s,r}^{(m,d)} 
  :=  (\tilde{H}_{m,d})_{n,k} \delta_{j,\ell} 
    + (\tilde{H}_{m,d})_{j,\ell}\delta_{n,k}.
\end{equation}
As before, we do this for all system and control Hamiltonians, stack the
resulting Liouville operators, and calculate the null space of the
resulting matrix.  Unstacking the null vectors again results in ASOs $\tilde{J}$ such that $\tilde{H}_{m,d}^T\tilde{J}+\tilde{J}
H_{m,d}=0$ for all $\tilde{H}_{m,d}$.  The ASOs
$\tilde{J}$ define the possible superpositions within the subspaces
spanned by the eigenvectors of the $J$ CSOs.  The eigenvalues of
the $\tilde{J}$ may help identify the Lie subalgebras \cite{Lieident,
completecontrol}.

\end{document}